\documentclass[aps,prl,twocolumn,amsmath,amssymb,floatfix,showpacs]{revtex4}
\usepackage{graphicx}% Include figure files
\usepackage{amsmath}
\usepackage{color}% add textcolor
\usepackage{epstopdf}% Allow .eps files
\usepackage{dcolumn}% Align table columns on decimal point
\usepackage{bm}% bold math
\usepackage{epsfig}

\begin{document}
\graphicspath{{Figures/}}
%\preprint{APS/123-QED}

\title{Non-equilibrium Pattern Modes Extracted from Experimental Data}
\author{Adam~C.\ Perkins, Roman~O.\ Grigoriev, and Michael~F.\ Schatz}
\affiliation{Center for Nonlinear Science and School of Physics, Georgia Institute of Technology, Atlanta, Georgia 30332, USA}
%\date{\today}
\begin{abstract}

%%%%% Abstract
We describe a method to extract from experimental data the important dynamical modes in spatio-temporal patterns in a system driven out of thermodynamic equilibrium. Using a novel optical technique for controlling fluid flow, we create an experimental ensemble of Rayleigh-B\'{e}nard convection patterns with nearby initial conditions close to the onset of secondary instability.  An analysis of the ensemble evolution reveals the spatial structure of the dominant modes of the system as well as the corresponding growth rates. The extracted modes are related to localized versions of instabilities found in the ideal unbounded system. The approach may prove useful in describing instability in experimental systems as a step toward prediction and control.

\end{abstract}

% Flow instabilities, Chaos in Fluid Dynamics, Transition to Turbulence

% Flow Control
%47.85.L-

\maketitle

%%%%% Introduction
Identification of instabilities plays a crucial role in our understanding and description of the the dynamics of many nonlinear physical, biological, and chemical systems driven out of equilibrium~\cite{cross_hohenberg}. In particular, quantitative description is essential for predicting and/or controlling the evolution of such systems, with weather prediction being a prime example. While linear stability analysis of global disturbances in an idealized, infinite system may provide a description of dynamics and pattern selection, this approach fails for imperfect patterns (e.g., far from onset) and in strongly confined systems. Despite recent numerical advances~\cite{egolf_nature, Scheel_Cross, Paul_2007} in computing the spatial structure and dynamics of localized disturbances in weakly chaotic patterns, no general approach has been developed for extracting such dynamical information directly from experimental measurements.

In this Letter, we present such an approach, illustrating how dynamical degrees of freedom can be extracted from experiments conducted on the prototypical Rayleigh-Benard convection (RBC) system. Specifically, we determine the spatial structure and evolution of the dominant dynamical degrees of freedom by analyzing the response of the system to an ensemble of localized perturbations about the straight roll state. In each case, we find the dynamics are dominated by a small number of spatially-localized modes. We also observe slowing down of the dynamics and thus quantify the distance to a particular instability boundary in terms of perturbation lifetimes. The spatial structure of the extracted modes is found to be consistent with the classification of secondary instabilities of spatially infinite perfect patterns.

%%%%%%
\begin{figure}[t]
\centering
\includegraphics[scale = 0.175]{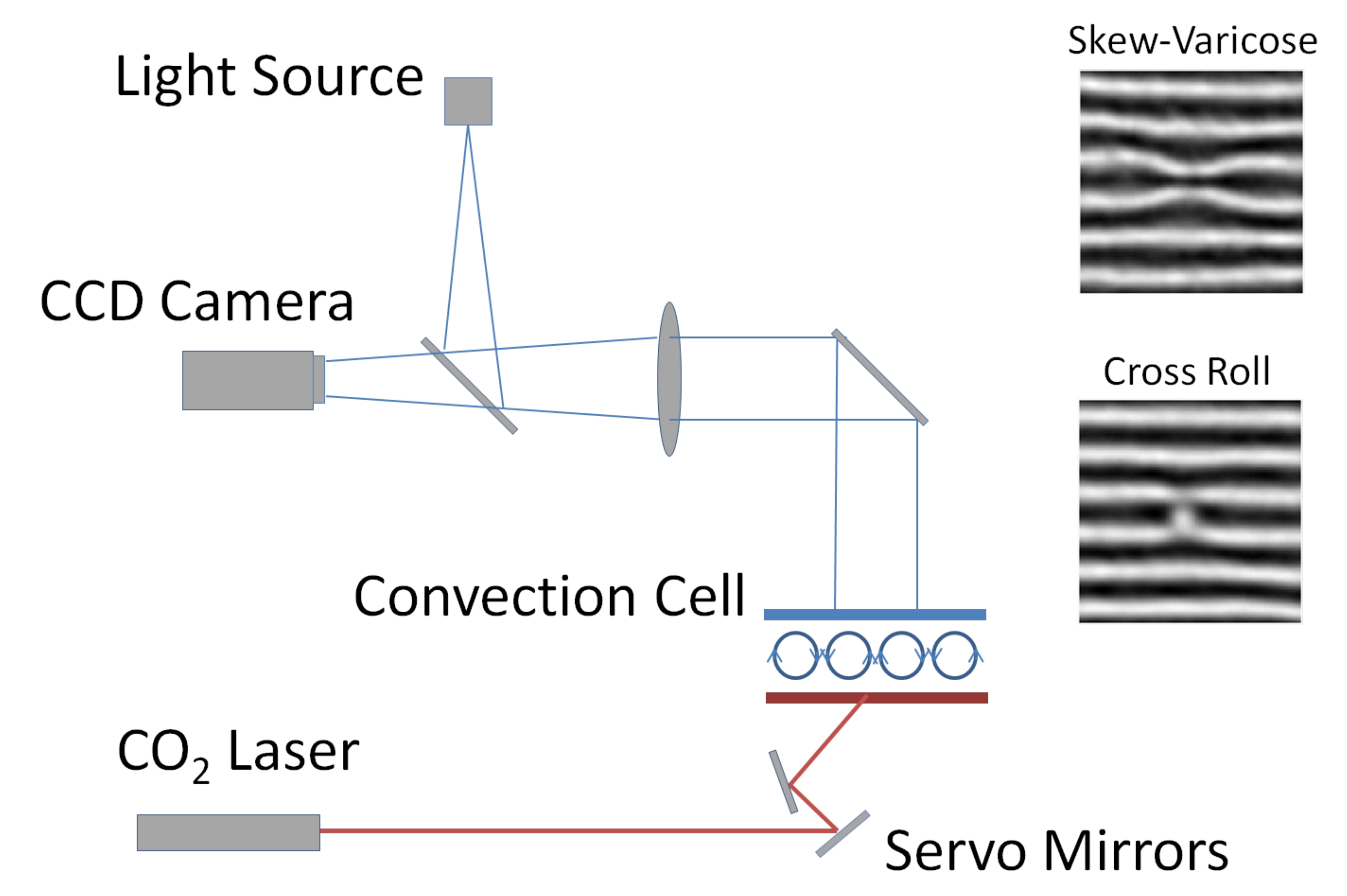}
\caption{\label{fig:schematic} Experimental setup includes a shadowgraph visualization as well as a system for optical actuation. Inset are observed responses to local perturbations at different parameter values. These disturbances represent the local version of the skew-varicose and cross-roll instabilities, respectively, of the unbounded system.}
\end{figure}
%%%%%%

%%%%% Experimental
The convection experiments were performed with a layer of sulfur hexafluoride (SF$_{6}$) gas of depth $d = 700 \pm 10 {\mu}m$ compressed to $14.5 \pm 0.1$ bar \cite{rbc_apparatus}. The gas layer was confined laterally to a 25 mm x 15 mm rectangular region by filter paper sidewalls chosen to match the fluid conductivity. The layer was bounded from above by a water-cooled sapphire window and from below by a carbon-disulfide (CS$_{2}$)-cooled zinc selenide (ZnSe) window; the temperatures of these windows were regulated to $\pm 0.05~^{\circ}$C. Experiments were performed with $\bar{T}=24.00~^{\circ}$C and temperature differences of $2.92$ and $4.80~^{\circ}$C. Convection was visualized using the shadowgraph method by illuminating from above and imaging the light reflected from the ZnSe surface at the bottom of the gas layer.

SF$_{6}$ is a greenhouse gas; we use this property to optically apply controlled thermal disturbances both to manipulate the global convective flow as well as impose localized perturbations. An IR beam from a CO$_{2}$ laser ($10.6$ ${\mu}m$ wavelength) is focused to approximately 200 ${\mu}m$ in diameter and strikes the gas layer from below (after passing through the IR transparent CS$_2$ coolant and ZnSe window). The beam is steered by two computer-controlled, gold-plated servo mirrors that rotate about orthogonal axes, providing the ability to direct the IR light toward any point over the cell domain. Fig.~\ref{fig:schematic} shows the experimental setup. The extinction length of the beam in the SF$_{6}$ is $<$ 10 ${\mu}m$~\cite{SF6_absorption}, less than 2\% of the cell depth, so the absorbed beam induces a highly localized heating that takes place very near the bottom of the gas layer.  Software developed in-house works with a commercial program (LaserShow2000) to synchronize laser power with mirror rotations. The strong absorption and rapid scanning results in the ability to manipulate the flow on a time scale much faster than the typical dynamical time scale (vertical thermal diffusion time of 2.7 s). This technique improves on previous attempts to manipulate convection patterns by optical actuation~\cite{manip_busse, schatz_hexagons}.

A pattern of straight rolls with wavenumber $q$ was established by sending pulses of laser light into the cell at desired hot roll locations. To minimize sidewall effects, the pattern was imposed in the central portion of the cell, 1-2 wavelengths from either of the sidewalls in the wavevector direction. This provides room for 7-9 interior pairs of straight rolls.

\begin{figure}[t]
\centering
\includegraphics[trim=0.2cm 6.4cm 0.5cm 7.0cm, clip=true, scale=0.39]{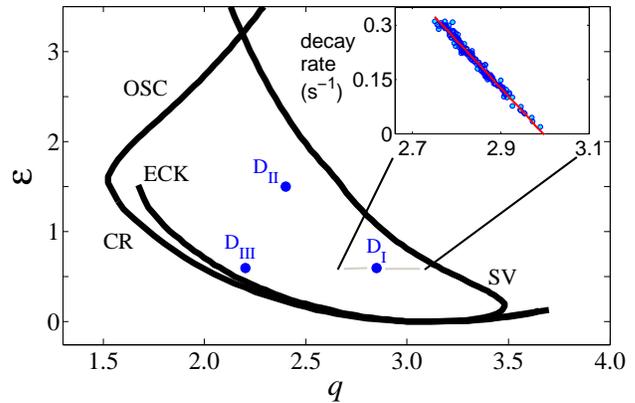}
\caption{\label{fig:Busse} Stability balloon at $Pr = 0.84$ showing various instabilities (CR = Cross Roll, ECK = Eckhaus, SV = Skew Varicose, OSC = Oscillatory) of the straight roll state along with the parameters at which the experimental data were taken for the three data sets.  $D_{I}$ = (0.60, 2.85), $D_{II}$ = (1.50, 2.40), and $D_{III}$ =  (0.60, 2.20). The inset plot shows the disturbance decay rate ($s^{-1}$) as a function of wavenumber $q$ at fixed $\epsilon = 0.60$.}
\end{figure}

Secondary instabilities of the straight roll state are predicted to define the boundaries of a stability (Busse) balloon in $({\epsilon},q)$ space at fixed Prandtl number; crossing one of these instability thresholds results in a re-organization of the pattern~\cite{busse}. Here, ${\epsilon} = (R - R_{c})/R_{c}$ is the reduced bifurcation parameter that measures the distance from onset of the primary instability of the purely conducting state. Fig.~\ref{fig:Busse} shows the Busse balloon for the spatially unbounded system at the conditions of our experiments ($Pr = 0.84$). The Busse balloon is conventionally calculated assuming an infinite system; we may expect the thresholds for the localized instabilities in a bounded system to take place at slightly different parameter values. Nonetheless, the balloon provides a useful reference as different areas in the parameter space are visited, and the dynamics of the dominant modes of the system can be expected to depend on the distances between a given $({\epsilon},q)$ point and the various instability boundaries. 

Pattern manipulation allows us to use both $\epsilon$ and $q$ as control parameters. The initial pattern wavenumber was chosen to be within the stable band. Following the inital imprinting, actuation is restricted to only the two outer-most rolls between which the straight roll pattern is confined.  The pattern wavenumber is adjusted by moving the positions of these outer rolls toward or away from one another; the interior pattern equilibrates on a timescale of approximately 10 $t_{v}$ for our domain size. A closed-loop feedback algorithm constantly analyzes images for deviations of the two boundary roll positions from desired locations and adjusts laser power accordingly.

Sufficiently close to a particular instability boundary, the spatial modes corresponding to that instability are weakly damped and can therefore be excited by small perturbations to the base state (straight rolls). Each perturbation takes the form of a brief (about 100 ms) well-localized laser pulse directed to a single location in the convection pattern. We estimate that the applied heating results in an initially axisymmetric disturbance of diameter $\sim d/2$.

%%%%%%
%%%%%

%%%%% Results and Discussion
We first probed the pattern response to perturbations for different $q$ near the high-wavenumber instability boundary, with fixed $\epsilon = 0.60$. Upon perturbation, the two hot rolls appear to bend toward one another and then relax back to their initial locations; a snapshot of a typical response is shown in the inset of  Fig.~\ref{fig:schematic}. The disturbance decay is slower at higher $q$, which suggests using the perturbation lifetime as a measure of distance from instability. The perturbation lifetime is measured most easily from the apparent local roll separation as a function of time. This signal typically displays a large spike immediately after the perturbation, followed by exponential decay to the original value; analysis is restricted to this period of linear decay.

We find that the disturbance lifetime increases linearly with the pattern wavenumber over an order of magnitude change in decay rate (see the inset in Fig.~\ref{fig:Busse}). The decay-rate distribution shows the expected slowing down of the dynamics near the instability and indicates the critical value $q_{c} = 3.01$ for the localized skew-varicose instability. Note that this value is slightly smaller than that of the global instability predicted from analysis of an infinite domain ($q_{c} = 3.15$), which reflects the effect of spatial localization. The low amount of scatter in the decay-rate plot illustrates the high degree of reproducibility of the imposed perturbations.

%%%%% Algorithm
In order to excite all dominant localized modes, perturbations were applied at a grid of equally-spaced locations across a wavelength of the pattern. All other perturbations are related to this set through the symmetries of the system (translational invariance in the direction perpendicular to the wavevector, periodicity in the direction of the wavevector), as long as the disturbances are sufficiently well-localized and not near the physical boundaries. Each set contains perturbations at 12 distinct locations; in all experiments, the spatial extent of the laser is of order $10\%$ of the pattern wavelength, so there is no benefit from a finer partition.

Shadowgraph images capture the full evolution of an imposed disturbance and thus contain snapshots of the composite structure of the excited modes over time. During the perturbation decay, there exists a period of time over which the dynamics of the disturbance can be described by a linear evolution operator; we seek to represent this evolution operator by computing its matrix elements in a subspace spanned by a set of slow modes extracted from the shadowgraph images. We first subtract the stationary straight roll pattern from all images after the perturbation. The images are then spatially windowed and Fourier filtered. A Karhunen-Lo\'{e}ve (KL) decomposition of difference images representing perturbations 2-5 seconds after each initial disturbance provides a set of basis modes. All perturbations are then projected onto the subspace spanned by these basis modes. Note that the typical implementation of the KL decomposition uses time-averaging~\cite{sirovich}, whereas we employ an ensemble average over different initial conditions. We limit our embedding dimension to a small (usually three or four) number of modes that capture 90\% of the power of the KL eigenvalue spectrum. 

Let us denote the disturbance following an initial perturbation ${\textbf{b}}^0$, expressed in a low-dimensional basis. After some time $T$, this state has evolved to ${\textbf{b}}^T$.  Then $U\textbf{b}^0 = \textbf{b}^T$, where $U$ is the evolution operator. Using an ensemble of initial conditions we define \[{B}^0 = \begin{bmatrix} {\textbf{b}}^0_1 & \vline & {\textbf{b}}^0_2 & \vline & \cdots & \vline & {\textbf{b}}^0_M \end{bmatrix}\]

and similarly, \[{B}^T = \begin{bmatrix} {\textbf{b}}^T_1 & \vline & {\textbf{b}}^T_2 & \vline & \cdots & \vline & {\textbf{b}}^T_M \end{bmatrix}\]

This gives the over-determined (least-squares) problem for the evolution operator $UB^0 = B^T$, which is solved by $U = B^T{(B^0)}^{-1}$, where the reciprocal of $B^0$ is taken to refer to the generalized inverse of the non-square matrix.

During linear decay, each system eigenmode decays at a characteristic rate, so we can also write $U \textbf{e}_{i} = \exp({\sigma{T}})\textbf{e}_{i}$, where $\textbf{e}_{i}$ are the eigenvectors and the eigenvalues ${\lambda}_{i}$ are related to the growth rates by ${\lambda}_{i} = \exp({\sigma_{i}{T}})$. %{e}^{\sigma_{i}{T}}$.

%%%%%%
\begin{figure}[h]
\centering
\begin{tabular}{c c}
\includegraphics[trim=6cm 9cm 5cm 9cm, clip=true, scale=0.40]{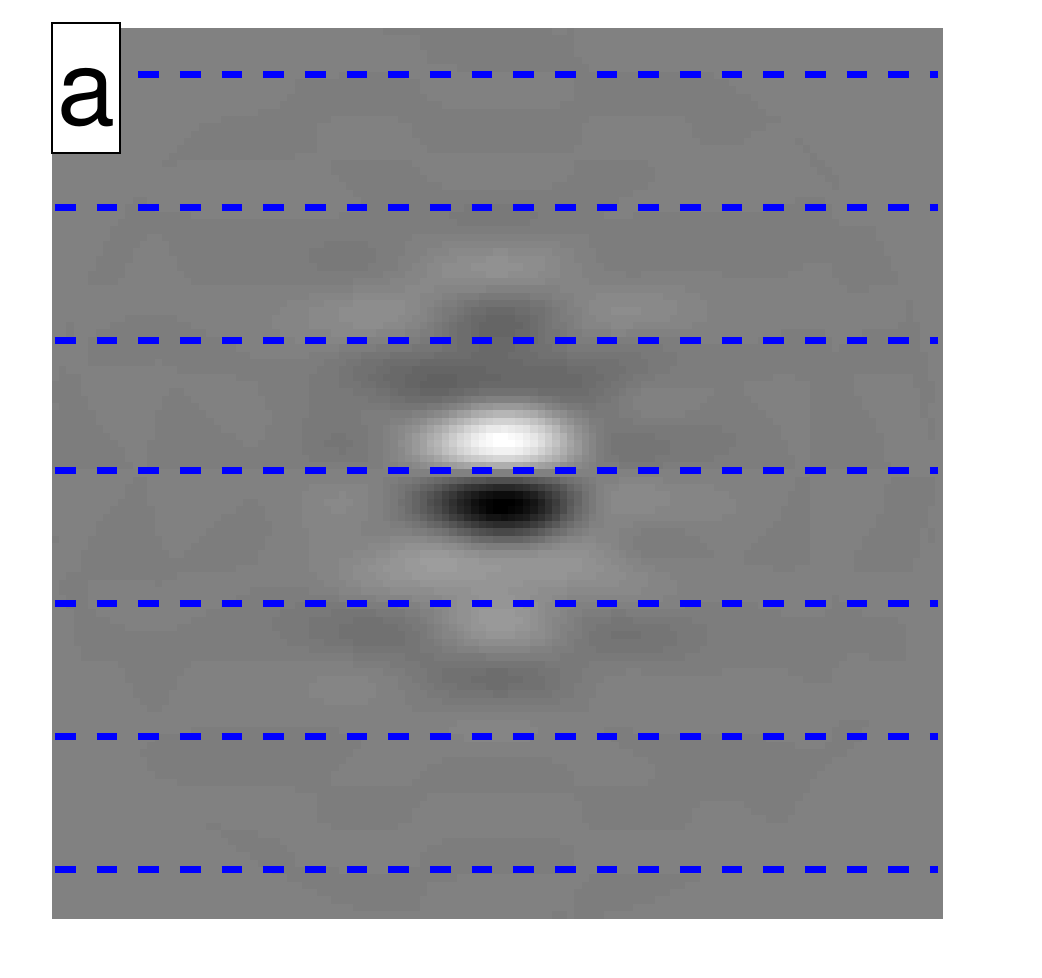} &
\includegraphics[trim=6cm 9cm 5cm 9cm, clip=true, scale=0.40]{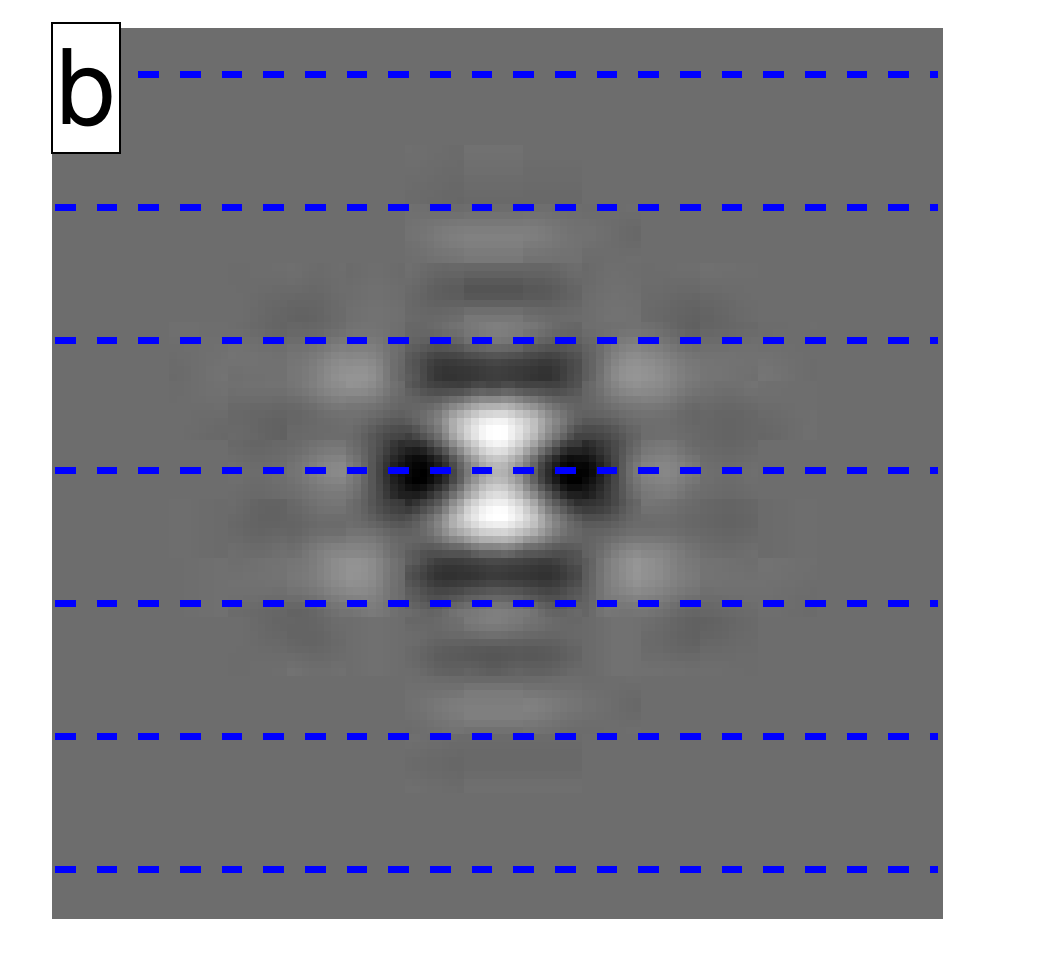} \\ 
\end{tabular}
\caption{\label{fig:DI} (a) The fundamental dominant and (b) sub-dominant mode extracted at $D_{I}$.  Dashed lines mark the approximate locations of the hot rolls of the underlying base state; cold rolls lie directly between the hot rolls.}
\end{figure}
%%%%%%

%%%%%
\begin{figure}[h]
\centering
\begin{tabular}{c c}
\includegraphics[trim=6cm 9cm 5cm 9cm, clip=true, scale=0.40]{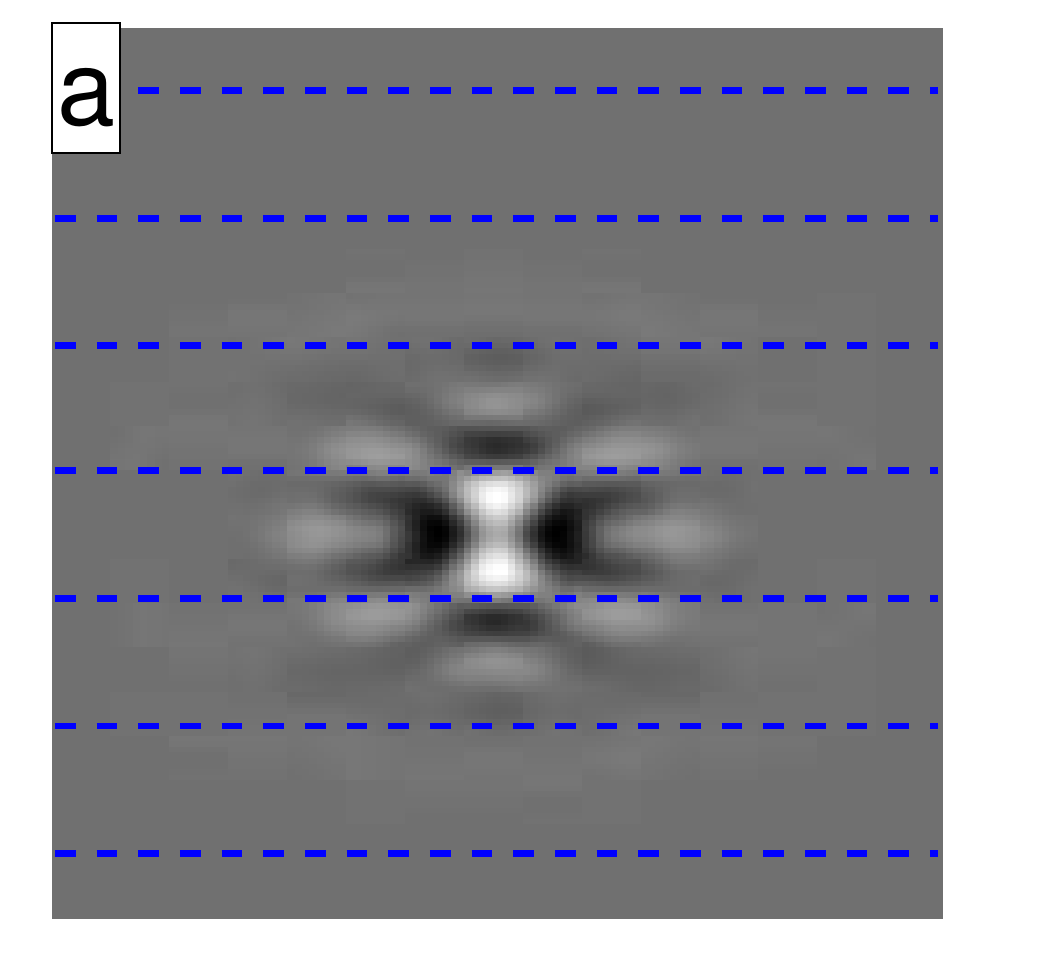} &
\includegraphics[trim=6cm 9cm 5cm 9cm, clip=true, scale=0.40]{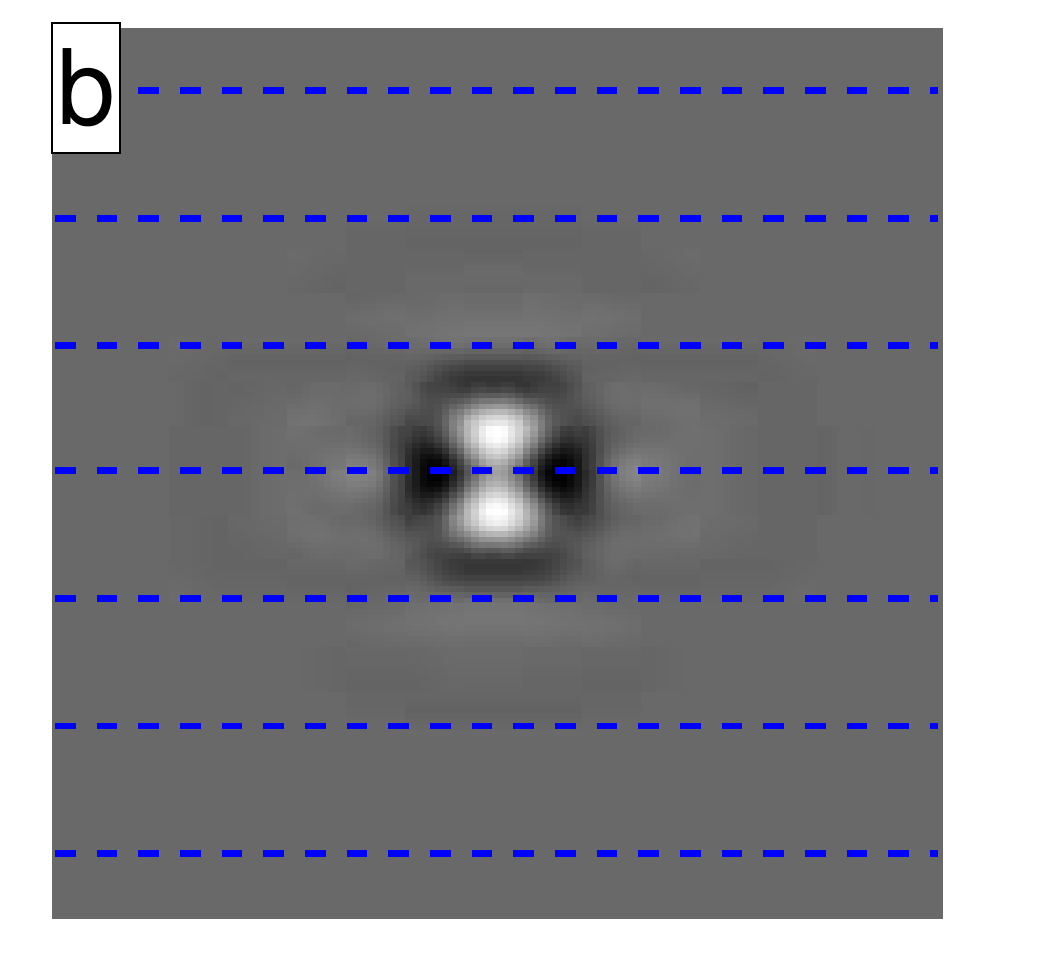} \\ 
\end{tabular}
\caption{\label{fig:DIII} (a) The fundamental dominant and (b) sub-dominant mode extracted at $D_{III}$.}
\end{figure}
%%%%%

The discrete translational symmetry of the straight roll pattern implies that a superposition of a well-localized eigenmode with a copy of itself translated over an integer number of wavelengths is an eigenmode with the same growth rate. Hence, there exist many possible representations of each eigenmode. Reduction of the dynamics using the symmetries of the base pattern is useful for both unbounded and bounded systems.

In particular, all disturbances can be decomposed in terms of symmetries related to a pair of hot and cold rolls.  By defining two symmetry planes, one at the center of a hot roll, the other at the center of an adjoining cold roll, we can define four symmetric versions of every initial disturbance, each even/odd about the hot/cold roll. Each of the corresponding four subspaces is invariant: disturbances retain their symmetry as they evolve.

The entire collection of initial and final conditions extracted from experiment was decomposed using these symmetries, producing four independent ensembles. All eigenmodes extracted from the four ensembles are eigenmodes of the system, but in the cases when multiple eigenmodes share an eigenvalue (growth rate) we eliminate redundant representations by computing the most spatially-localized eigenmode structure.

We estimate the uncertainty in the growth rates to be less than 10\%. This allows one to group all extracted modes and define the fundamental mode as the most-localized structure with a particular growth rate. The fundamental modes were computed by minimizing the p-norm ($p < 2$) among linear combinations of all modes (and their translated copies) in each group. The symmetries imply that all extracted modes in each group can be represented as linear superpositions of the fundamental mode along with its translated and/or reflected copies. We verified that this is indeed the case as such representation was accurate, with mutual projection $> 0.94$.

%%%%% Discussion
The first ensemble was produced at the point $D_{I}$ of the parameter space (see Fig.~\ref{fig:Busse}). The two dominant modes are shown in Fig.~\ref{fig:DI}. As the lifetime measurements indicate, the least-stable mode ($\sigma_{1}$ = -0.13 $s^{-1}$) tended to be excited from perturbations between two hot rolls (at the location of a cold roll).  Note, however, that while the structures excited from these disturbances are even about the cold roll, the most-localized representation of this mode does not obey that symmetry. The sub-dominant mode ($\sigma_{2}$ = -0.70 $s^{-1}$) tended to be excited from perturbations directly at the location of a hot roll.

A second set of perturbations at high $q$ was performed at $D_{II}$, with $\epsilon$ increased relative to $D_{I}$. Again, two modes are extracted; $\sigma_{1}$  = -0.15 $s^{-1}$ and $\sigma_{2}$ = -0.55 $s^{-1}$. Mutual projections of the modes extracted at $D_{I}$ and $D_{II}$ indicate that the spatial structure of the two dominant modes remains unchanged (after scaling by the wavelength) between these two locations. There is also, in both cases, a large separation between the two growth rates. This suggests we can identify the dominant mode (Fig.~\ref{fig:DI}a) as the one representing the secondary instability at the high wavenumber boundary. Its structure is consistent with a skew-varicose type instability.

We also created an ensemble of perturbations to a low-wavenumber pattern, at $D_{III}$. As in the other two cases, two modes were extracted, with growth rates $\sigma_{1}$ = -0.20 $s^{-1}$ and $\sigma_{2}$ = -0.27 $s^{-1}$. As seen in Fig.~\ref{fig:DIII}, the spatial structure of the dominant mode does not resemble any of the previously extracted modes, while the sub-dominant mode resembles the sub-dominant mode extracted at both $D_{I}$ and $D_{II}$. The closeness of the growth rates is consistent with the existence of two low-wavenumber instability types of the unbounded system that occur at very nearly the same parameter values, namely, the Eckhaus and cross-roll instabilities (see the Busse balloon in Fig.~\ref{fig:Busse}). The dominant mode is again excited from perturbations to a cold roll; this is not surprising, as heating of cold rolls tends to reduce the amplitude of the saturated state. We find experimentally that sufficiently strong perturbations of this kind result in the growth of rolls perpendicular to the base pattern. We therefore identify the dominant low-wavenumber mode with the localized cross-roll instability and the sub-dominant mode with the localized Eckhaus instability.

%%%%% Conclusions
Further experiments are needed to explore the applicability of this approach to states exhibiting more complex dynamics. One such state, occurring in gas convection experiments with $Pr \approx 1$, is the spatiotemporally chaotic state known as Spiral Defect Chaos (SDC)~\cite{morris_SDC}. While there exists a bistability between stationary straight rolls and SDC~\cite{bistability} over the parameter range of the experiments reported here, localized instability in the straight roll pattern introduces defects which tend to lead to a disordered pattern, thus providing a mechanism for the transition to chaotic behavior. Additionally, it was determined in a numerical study~\cite{egolf_nature} that the chaotic dynamics of SDC are largely driven by the creation/annihilation of defects occuring in straight roll regions of the pattern. We expect, therefore, that spatially-localized modes are dynamically-important in both the transition to and the driving of chaotic behavior, suggesting a natural extension of our experimental approach to investigations of more complex convection patterns. Moreoever, the outlined procedure is general enough to be used in a variety of other dynamical systems, so long as an appropriate means of system actuation can be developed. In addition to being of fundamental interest and of use in increasing predictive power, knowledge of the modes of instability could be particularly advantageous in system control, where small, controlled perturbations could be used to guide system dynamics~\cite{ott_control}.

%%%%% Acknowledgements
\begin{acknowledgments}
The support of this project by the National Science Foundation through Grant AGS-0434193 is gratefully acknowledged.  The authors would like to thank Gabriel Seiden for suggesting the use of infrared light to actuate convective flow of SF$_{6}$. We would also like to thank Werner Pesch for helpful comments and the code used to produce the Busse balloon.
\end{acknowledgments}   

%%%%% Bibliography

\end{document}